\documentstyle[sprocl]{article}

\begin{document}

\title{Test particle description of transport processes for states with 
a continuous mass spectrum
} 

\author{Stefan Leupold}

\address{Institut f\"ur Theoretische Physik, Justus-Liebig-Universit\"at
Giessen,\\
D-35392 Giessen, Germany}


\maketitle\abstracts{%
Aiming at a description of transport processes where the dynamically generated width 
of the states is potentially large a transport equation beyond the quasiparticle 
approximation is derived in first order gradient expansion. An effective particle
number is identified which is exactly conserved by the coarse grained transport 
equation. Using a test particle ansatz for this conserved 
quantity allows to rewrite the transport equation
into equations of motion for test particles. The two-body collision terms are formulated
in terms of the test particles which gain non-trivial renormalization factors due to the 
coarse graining process.}

\section{Motivation}

This talk is based on my recent work \cite{SL99}. Due to limited space I refer to this
work \cite{SL99} concerning more details and references. 

For the description of heavy-ion collisions the following Boltzmann-type equation
is commonly used:
\begin{equation}
  \label{eq:boltz}
\left(\partial_t + {\vec p \over m}\vec\nabla_x - \vec\nabla_x U(t,\vec x) \vec \nabla_p
\right) 
f(t,\vec x,\vec p) = I_{\rm coll}
\end{equation}
where the l.h.s.~contains the drift and the Vlasov term and the r.h.s.~incorporates
the collisions. The central quantity here is the {\em on-shell} phase space distribution
$f$. ``On-shell'' simply means that the energy is not an additional free
variable on which $f$ might depend but is instead given by the appropriate mass-shell
condition. Therefore $f$ depends only on the three-momentum (and the space-time 
variables). I will discuss in a moment why this might be an unpleasant restriction
for various processes relevant for the description of heavy-ion reactions.
Before I come to that point I want to discuss briefly how eq.~(\ref{eq:boltz}) can be
solved: One usually involves a test particle ansatz
\begin{equation}
  \label{eq:testp1}
f(t,\vec x,\vec p) \sim \sum\limits_i \delta^{(3)}(\vec x - \vec x_i(t)) \, 
\delta^{(3)}(\vec p - \vec p_i(t)) 
\end{equation}
and deduces from (\ref{eq:boltz}) the collisions of these test particles and the
evolution of their coordinates $\vec x_i$, $\vec p_i$ between collisions. In what
follows I will concentrate on the evolution part. The equations of motion for the test
particle coordinates {\em between collisions} are obtained by inserting 
(\ref{eq:testp1}) in (\ref{eq:boltz}) and replacing the collision term by zero. In this
way one obtains the equation
\[
\sum\limits_i \left( \left[ -\dot{\vec x_i} + {\vec p_i \over m} \right] \vec\nabla_x
+ \left[ -\dot{\vec p_i} - \vec \nabla_{x_i} U(t,\vec x_i) \right] \vec\nabla_p \right)
\delta^{(3)}(\vec x - \vec x_i) \,\delta^{(3)}(\vec p - \vec p_i) = 0  \,.
\]
This equation can be fulfilled by demanding that both expressions in square brackets
have to vanish independently for every test particle $i$. In this way one gets the
familiar Newtonian equations for the test particle coordinates
\[
\dot{\vec x_i} = {\vec p_i \over m} \qquad , \qquad 
\dot{\vec p_i} = - \vec \nabla_{x_i} U(t,\vec x_i)  \,.
\]
From a practical point of view one expects to get a sufficiently correct expression for
the phase space density if the number of test particles is reasonably large. 
This traditional transport theoretical framework is capable to describe states with 
energies close to or at the mass-shell point. For heavy-ion collisions at sufficiently 
high energy, however, broad resonances come into play with decay widths of the order of 
the typical kinetic energies. In addition, also the width due to collisional broadening
might become comparably large. In this case the quasi-particle approximation which enters
the field theoretical derivation of the Boltzmann equation (\ref{eq:boltz})
becomes questionable and one has to find extensions of that model. The new central
quantity of interest will be the Wigner transform (see below) of the two-point function
denoted by $S^<(t,\vec x;p_0,\vec p)$. Note that this quantity depends explicitly on
the energy and not only on the three-momentum. It is the purpose of the next section
to derive a transport equation for $S^<$ {\em without} resorting to the quasi-particle
approximation. Then I will discuss how the obtained transport equation can be solved by
an appropriate test particle ansatz. One of the questions will be for which quantity
a test particle ansatz is justified. From the discussion outlined above it is already now
apparent that in such an extended test particle ansatz the energy has to be treated
as an independent variable. Thus, this ansatz for a so far unspecified quantity has the 
form
\begin{equation}
  \label{eq:from}
\tilde S^<(t,\vec x;p) \sim 
\sum\limits_i \delta^{(3)}(\vec x - \vec x_i(t)) \, \delta(p_0 - E_i(t)) \,
\delta^{(3)}(\vec p - \vec p_i(t)) \,.
\end{equation}

\section{From the Kadanoff-Baym equations to the transport equation}

It is obviously appropriate for a talk at this workshop to start out from the 
Kadanoff-Baym equations:
\begin{eqnarray}
\label{eq:KBkl}
\left(i {\partial \over \partial t_1} + {\Delta_1 \over 2m}\right) D^<(1,1') &=&
\int \!\! d\bar 1 
\left[
    \Sigma^{\rm ret}(1,\bar 1) \, D^<(\bar 1, 1') 
  + \Sigma^<(1,\bar 1) \, D^{\rm av}(\bar 1, 1')
\right]  
\phantom{mj} \\
\label{eq:KBgr} 
\left(i {\partial \over \partial t_1} + {\Delta_1 \over 2m}\right) D^>(1,1') &=&
\int \!\! d\bar 1 
\left[ 
    \Sigma^{\rm ret}(1,\bar 1) \, D^>(\bar 1, 1') 
  + \Sigma^>(1,\bar 1) \, D^{\rm av}(\bar 1, 1')
\right]  
\\
\label{eq:KBret}
\left(i {\partial \over \partial t_1} + {\Delta_1 \over 2m}\right) 
D^{\rm ret}(1,1') &=&
\delta^{(4)}(1,1') 
+ \int \!\! d\bar 1 \, \Sigma^{\rm ret}(1,\bar 1) \, D^{\rm ret}(\bar 1,1')
\end{eqnarray}
with
\begin{eqnarray*}
i D^< (x,y) &=& \pm \langle \psi^\dagger(y) \,\psi(x) \rangle \,, \qquad  
i D^> (x,y) = \phantom{\pm} \langle \psi(x) \,\psi^\dagger (y) \rangle \,,
\\ 
D^{\rm ret}(x,y) &=& \Theta(x_0-y_0) \left[ D^>(x,y) -D^<(x,y) \right] \,, \\ 
D^{\rm av}(x,y) &=& \Theta(y_0-x_0) \left[ D^<(x,y) -D^>(x,y) \right]  \,.
\end{eqnarray*}
The plus/minus sign refers to bosons and fermions, respectively. 
Actually all information of this non-relativistic system is already contained in the
first two equations (\ref{eq:KBkl}) and (\ref{eq:KBgr}). In what follows it is more 
convenient however to additionally deal with the equation (\ref{eq:KBret}) for the 
retarded propagator. As usual for the derivation of kinetic equations from the 
underlying field theoretical equations I assume now that, concerning the transport 
phenomena one wants to describe, the dependence of all two-point
functions and self-energies on their respective (macroscopic) center-of-mass variables 
$X=(x+y)/2$ is much weaker than the dependence on the (microscopic) difference variable
$u=x-y$. 
In this case it is useful to perform a Fourier
transformation of all quantities with respect to their respective difference variable
(Wigner transformation)
\[
\bar D^<(X,p) = \int \!\! d^4\!u \, e^{ipu} D^<(X+u/2,X-u/2)  \,,
\]
expand in gradients with respect to $X$ and neglect all terms which contain more than
one derivative in $X$. I also introduce the generalized Poisson bracket 
\[
[A,B] = \partial_{X_0} A \,\partial_{p_0} B - \partial_{p_0} \,A \partial_{X_0} B 
- \vec\nabla_X A \,\vec\nabla_p B + \vec\nabla_p A \,\vec\nabla_X B   \,.
\]
By the procedure outlined above the Kadanoff-Baym equations 
(\ref{eq:KBkl}-\ref{eq:KBret}) turn into three (gradient expanded) complex-valued 
equations. These 
equations can be separated into their real and imaginary parts by introducing the
following real-valued quantities
\begin{eqnarray*}
S^<(X,p) &=& \pm i \bar D^<(X,p) \,,  \\
{\cal A}(X,p) &=& -2 {\rm Im}\bar D^{\rm ret}(X,p) 
= i[\bar D^>(X,p) - \bar D^<(X,p)]  \,.
\end{eqnarray*}
From the Kadanoff-Baym equations one gets by gradient
expansion the following set of six (real-valued) equations:
\begin{eqnarray}
  \label{eq:anasless}
\left(p_0 - {\vec p^2 \over 2m} - {\rm Re}\bar \Sigma^{\rm ret} \right) S^< &=&
\pm i \bar\Sigma^< \,{\rm Re}\bar D^{\rm ret}
-{1 \over 4}[\Gamma,\, S^<] + {1 \over 4}[\pm i\bar\Sigma^<,\,{\cal A}] \,, \phantom{mm}
\\ 
  \label{eq:gradsless}
[p_0 - {\vec p^2 \over 2m} - {\rm Re}\bar \Sigma^{\rm ret} ,\, S^<] &=&
\Gamma \, S^< \mp i \bar\Sigma^< \,{\cal A}
+ [\pm i \bar\Sigma^<, \, {\rm Re}\bar D^{\rm ret}]   \,,
\\ 
  \label{eq:anaspec}
\left(p_0 - {\vec p^2 \over 2m} - {\rm Re}\bar \Sigma^{\rm ret}  \right) {\cal A} &=& 
\Gamma \, {\rm Re}\bar D^{\rm ret}   \,,
\\ 
  \label{eq:gradspec}
[p_0 - {\vec p^2 \over 2m} - {\rm Re}\bar \Sigma^{\rm ret} ,\, {\cal A}] &=&
[\Gamma, \, {\rm Re}\bar D^{\rm ret}]  \,,
\\
  \label{eq:anare}
\left(p_0 - {\vec p^2 \over 2m} - {\rm Re}\bar \Sigma^{\rm ret} \right) 
{\rm Re}\bar D^{\rm ret} &=& 
1 -{1 \over 4} \Gamma \, {\cal A}    \,,
\\  
  \label{eq:gradre}
[p_0 - {\vec p^2 \over 2m} - {\rm Re}\bar \Sigma^{\rm ret} ,\, {\rm Re}\bar D^{\rm ret}] 
&=& -{1 \over 4} [\Gamma, \, {\cal A}]   
\end{eqnarray}
where I have introduced the width $
\Gamma = i(\bar\Sigma^>- \bar\Sigma^<) = 
i(\bar \Sigma^{\rm ret}-\bar \Sigma^{\rm av})$.
In the end I am aiming at decoupled equations for the generalized phase space 
distribution $S^<$, the spectral function ${\cal A}$, and the real part of the 
retarded propagator ${\rm Re}\bar D^{\rm ret}$. Obviously one has twice as much 
equations as quantities which one wants to determine. Some of these equations must be 
redundant. From the purely algebraic equations (\ref{eq:anaspec},\ref{eq:anare}) one
immediately obtains 
\begin{eqnarray}
  \label{eq:solspec}
{\cal A}(X,p) &=& { \Gamma(X,p) \over 
\left( 
p_0 - {\vec p^2 \over 2m}-{\rm Re}\bar \Sigma^{\rm ret}(X,p)
\right)^2
+ {1 \over 4} \Gamma^2(X,p)  }        \,,    \\
  \label{eq:solred}
{\rm Re}\bar D^{\rm ret}(X,p) &=& 
{ p_0 - {\vec p^2 \over 2m}-{\rm Re}\bar \Sigma^{\rm ret}(X,p)  \over 
\left( 
p_0 - {\vec p^2 \over 2m}-{\rm Re}\bar \Sigma^{\rm ret}(X,p)
\right)^2
+ {1 \over 4} \Gamma^2(X,p)  }    \,.
\end{eqnarray}
These expressions automatically solve also eqs.~(\ref{eq:gradspec},\ref{eq:gradre}).
In the quasi-particle regime, 
i.e.~for $\Gamma \to 0$, the remaining two equations 
(\ref{eq:anasless},\ref{eq:gradsless}) turn into the
mass-shell constraint and the transport equation, respectively. For our case of 
arbitrary width $\Gamma$, however, these
two equations contain the same information provided one realizes by inspection of 
(\ref{eq:gradsless}) that the combination 
$\pm i \bar\Sigma^< - {\Gamma \over {\cal A} } S^<$ is effectively a first order 
gradient.
Since all equations are exact up to (including) first order gradients the following
replacement on the r.h.s.~of (\ref{eq:anasless},\ref{eq:gradsless}) is
allowed:
\begin{equation}
  \label{eq:repl}
[\pm i\bar\Sigma^<,\, \dots] 
\approx [{\Gamma \over {\cal A} } S^<, \, \dots ]   \,.
\end{equation}
After this replacement (\ref{eq:anasless}) and (\ref{eq:gradsless}) become identical.
Using (\ref{eq:solspec},\ref{eq:solred}) one finally gets the desired transport equation
for $S^<$:
\begin{eqnarray}
\lefteqn{{1 \over 2} \Gamma {\cal A} \,
[p_0 - {\vec p^2 \over 2m}-{\rm Re}\bar \Sigma^{\rm ret}, \, S^<] 
- {1 \over 2} {\cal A} \,
[ \Gamma, \, (p_0 - {\vec p^2 \over 2m}-{\rm Re}\bar \Sigma^{\rm ret})S^<] } 
\nonumber \\ 
&& \phantom{mmmmmmmmmmmmm} = \Gamma \, S^< \mp i \bar\Sigma^< {\cal A}  
\left( = i \bar\Sigma^> S^< \mp i \bar\Sigma^< S^>  \right)  \,.  
\label{eq:fintrans}
\end{eqnarray}
Before I will point out how this equation can be solved by a test particle ansatz I 
seemingly change my topic and discuss for which quantities a test particle ansatz
makes sense and for which quantities it does not.

\section{Test particles I: A simple example}

Imagine the following situation:
One has a theory describing the evolution of a quantity $f(t,p)$. This
theory, however, is too complicated to solve it exactly. (This is presumably easy to
imagine.) The only thing one knows about that theory is that it conserves a particle 
number $N(t) = \int\!\! dp \, f(t,p)$.
By an approximation scheme one obtains from the exact theory the following approximate
(transport) equation (which resembles a Vlasov equation):
\begin{equation}
  \label{eq:simp2}
{\partial \over \partial t} \left[ ( 1 - \kappa(t,p)) f(t,p) \right] + 
F(t) \, {\partial \over \partial p} f(t,p) = 0   
\end{equation}
with a force term $F(t)$ and a renormalization $\kappa$ which both are not further 
specified. Obviously this approximate (transport) equation does {\it not} conserve
the particle number $N$ but instead the quantity 
$\tilde N(t) = \int dp \, (1-\kappa(t,p)) f(t,p)$.
I know discuss two approaches to solve (\ref{eq:simp2}),
namely test particle ans\"atze for $f$ and for $\tilde f = (1-\kappa)f$, respectively: 
In general, a test particle representation for any of the two quantities is given by
\begin{equation}
  \label{eq:simp4}
\left.
  \begin{array}{c}
f \\ \mbox{or} \\ \tilde f
  \end{array}
\right\} =
{1 \over L} \sum\limits_{i=1}^M \delta(p-p_i(t))
\end{equation}
where $L$ is a normalization constant and $M$ the number of test particles. 
Obviously the momentum integral
over (\ref{eq:simp4}) is conserved and given by $M/L$. Therefore, a test particle ansatz
for $f$ to solve (\ref{eq:simp2}) seems to be inappropriate since the momentum integral
over $f$ (which yields $N$) is not conserved by (\ref{eq:simp2}). 
Nonetheless, it is  instructive to figure out which equations one gets with a test
particle ansatz for $f$. In this case one finds
\[
0 = \sum\limits_i  \left\{ - {d \kappa(t,p_i(t)) \over dt} 
+ \left[\vphantom{\int}
F(t)-[1-\kappa(t,p_i(t))]\dot p_i(t)\right] {\partial \over \partial p} \right\}  
\delta(p-p_i(t))
\,.  
\]
This yields {\em two} equations of motion for {\em one} quantity:
\begin{eqnarray*}
{d \kappa \over dt} =
{\partial \kappa \over \partial t} + \dot p_i {\partial \kappa \over \partial p_i} = 0
\quad \Rightarrow \quad \dot p_i &=& 
-{ {\partial \kappa \over \partial t} \over {\partial \kappa \over \partial p_i} }  \,,
\\
F = \dot p_i (1 - \kappa) \quad \Rightarrow \quad \dot p_i &=& { F \over 1 - \kappa} \,,
\end{eqnarray*}
i.e.~an in general overdetermined system. This is the consequence of the fact that
the integral over $f$, i.e.~$N$, is not conserved by the transport equation 
(\ref{eq:simp2}). Instead a test particle 
ansatz for $\tilde f= (1-\kappa)f$ yields
\begin{eqnarray*}
0 &=& \sum\limits_i  \left[\vphantom{\int}
{F(t) \over 1 - \kappa(t,p_i(t)) } - \dot p_i(t) \right] 
{\partial \over \partial p} \delta(p-p_i(t))  \,,
\end{eqnarray*}
i.e.~{\em one} equation of motion
\[
\dot p_i(t) = { F(t) \over 1 - \kappa(t,p_i)}   \,.
\]
Obviously this is a very intricate situation: by construction $N$ is conserved by the 
exact theory but not by the approximate transport equation. Therefore a test particle
ansatz for $f$ does not make sense. Thus I propose the following strategy: 
Solve the transport equation by a test particle ansatz for $\tilde f$ and calculate
from the solution the expression $dN/dt$. Since this expression should vanish in the
exact theory it provides a test for the accuracy of the approximation scheme which
has led to the transport equation (\ref{eq:simp2}). The lesson from this simple
example is the following: To solve a transport equation by a test particle ansatz one 
has to find a quantity which is {\em exactly} conserved by the approximate transport 
equation. It may appear that this quantity does not coincide with the one which is
exactly conserved by the exact (quantum field) theory.

\section{Test particles II: From the transport equation to the equations 
of motion for test particles}

For the full quantum field theory given by the Kadanoff-Baym equations 
(\ref{eq:KBkl}-\ref{eq:KBret}) the corresponding particle number is given by
\[
N(t) = \int  {d^3x \, d^4p \over (2\pi)^4} S^<(t,\vec x;p)  \,.
\]
For appropriately chosen self-energies (``$\Phi$-derivable'') this quantity is conserved.
I will restrict myself to the commonly used two-body collision terms (Born terms) which
indeed conserve $N$. Here e.g.~the out-rate is given by
\begin{eqnarray}
i \bar \Sigma^>(X,p) &=& 
\int\!\! {d^4\!p_1 \over (2\pi)^4}{d^4\!p_2 \over (2\pi)^4}{d^4\!p_3 \over (2\pi)^4} \,
(2\pi)^4 \delta^{(4)}(p+p_1-p_2-p_3) 
\nonumber \\
  \label{eq:born1}
&& \times {1 \over 2} 
\left(\bar v(\vec p -\vec p_2) \pm \bar v(\vec p -\vec p_3) \right)^2
S^<(X,p_1) \, S^>(X,p_2) \, S^>(X,p_3)  \,.  \phantom{mm}
\end{eqnarray}
If the quantity $N$ was also conserved by the transport equation (\ref{eq:fintrans}) 
a test particle ansatz for $S^<$ would make sense. However, one gets
\begin{eqnarray}
  \label{eq:notcons}
{d \over dt} N(t) 
= {d \over dt} 
\int\!\! d^3\!x \int\!\!{d^4\!p \over (2\pi)^4} S^< K
+ \int\!\! d^3\!x \int\!\!{d^4\!p \over (2\pi)^4} {2 \over \Gamma {\cal A}}
\left( \pm i \bar\Sigma^< S^> - i \bar\Sigma^> S^< \right)  \phantom{m}
\end{eqnarray}
with
\[
K(X,p) = {\partial {\rm Re}\bar \Sigma^{\rm ret}(X,p) \over \partial p_0} +
{p_0 - {\vec p^2 \over 2m}-{\rm Re}\bar \Sigma^{\rm ret}(X,p)  \over \Gamma(X,p) }
{\partial \Gamma(X,p) \over \partial p_0}   \,.
\]
I have not found any reason why the r.h.s.~of (\ref{eq:notcons}) should vanish and I 
strongly
conjecture that it does not. (Actually it is hard to {\em prove} that a quantity is
{\em not} conserved.) This conjecture is supported by the finding that a closely related
quantity, namely
\[
\tilde N(t) = \int  {d^3x \, d^4p \over (2\pi)^4} \tilde S^<(t,\vec x;p)  
\]
with $\tilde S^< = {1 \over 2} \Gamma {\cal A} S^< \, (1-K) $
is conserved by the transport equation (\ref{eq:fintrans}):
\[
{d \over dt} \tilde N(t) = 
\int\!\! d^3\!x \int\!\!{d^4\!p \over (2\pi)^4} 
\left( \pm i \bar\Sigma^< S^> - i \bar\Sigma^> S^< \right)= 0  
\]
where the last equation holds for the Born approximation to the self-energies.
I think it is hard to conceive that both $N$ and $\tilde N$ are exactly conserved
by the transport equation. Assuming that $N$ is not conserved a test
particle ansatz for $S^<$ is inappropriate. Instead, from the test particle 
ansatz (\ref{eq:from}) for $\tilde S^<$ which is the density corresponding to the 
conserved {\em effective} particle number $\tilde N$
one deduces the equations of motion for test particles {\it between two collisions}:
\begin{eqnarray*}
\dot E_i &=& 
{1 \over 1-K} \left(
\partial_t {\rm Re}\bar \Sigma^{\rm ret}
+ {\Delta E_i   \over \Gamma } 
\partial_t \Gamma 
\right)   \,,
\\ 
\dot{\vec x_i} &=&
{1 \over 1-K} \left(
{\vec p_i \over m} + \vec\nabla_{p_i} {\rm Re}\bar \Sigma^{\rm ret} +
{\Delta E_i   \over \Gamma } 
\vec\nabla_{p_i} \Gamma
\right)   \,,
\\ 
\dot{\vec p_i} &=&
- {1 \over 1-K} \left(
\vec\nabla_{x_i} {\rm Re}\bar \Sigma^{\rm ret} +
{\Delta E_i   \over \Gamma } 
\vec\nabla_{x_i} \Gamma
\right)   
\end{eqnarray*}
with the off-shellness $\Delta E_i(t,\vec x_i;E_i,\vec p_i) = 
E_i - { \vec p_i^2 \over 2m} - 
{\rm Re}\bar \Sigma^{\rm ret} (t,\vec x_i;E_i,\vec p_i) $.
The obviously new ingredient as compared to the traditional quasi-particle approximation
is this off-shellness. Its evolution in time can be obtained from the equations of 
motion given above to be
\begin{equation}
  \label{eq:tdoff}
{d \over dt} \Delta E_i = {\Delta E_i \over \Gamma} {d \over dt}\Gamma   \,.
\end{equation}
One apparent merit of this equation is the fact that test particles automatically 
get back to their mass-shell if the width goes to zero. I note in passing that a
crucial step to obtain (\ref{eq:tdoff}) consists in the replacement (\ref{eq:repl}).
Finally I want to comment on
the Born terms which in a test particle picture describe the collisions between these
test particles. Obviously e.g.~the out-rate (\ref{eq:born1}) has to be rewritten in
terms of test particle distributions:
\begin{eqnarray*}
\lefteqn{i \bar \Sigma^>(X,p) =} \nonumber \\ &&
\int\!\! {d^4p_1 \over (2\pi)^4}{d^4p_2 \over (2\pi)^4}{d^4p_3 \over (2\pi)^4} \,
(2\pi)^4 \delta^{(4)}(p+p_1-p_2-p_3) 
{1 \over 2} 
\left(\bar v(\vec p -\vec p_2) \pm \bar v(\vec p -\vec p_3) \right)^2  \nonumber \\ && 
\times
{\tilde S^<_1 \over {1\over 2}\Gamma_1\,{\cal A}_1\, (1-K_1)}
\, \left(
{\cal A}_2 \pm {\tilde S^<_2 \over {1\over 2}\Gamma_2\,{\cal A}_2 \, (1-K_2)} 
\right)
\, \left(
{\cal A}_3 \pm {\tilde S^<_3 \over {1\over 2}\Gamma_3\,{\cal A}_3 \, (1-K_3)} 
\right)  
\end{eqnarray*}
with $\tilde S^<_j = \tilde S^<(X,p_j)$ etc. The fact that $\tilde S^<$ instead
of $S^<$ is represented by test particles causes rather non-trivial modifications
for the collision integrals. In turn, disregarding these modifications amounts to 
solving a transport equation which might be rather different from the one which 
one actually wants to solve. 


\section{Test particles III: Comparison of the test particle method and 
Klimontovich's approach to kinetic theory}

Before summarizing I want to comment on the strong formal similarity of the test 
particle ansatz (\ref{eq:testp1}) to the microscopic phase space density 
\begin{equation}
  \label{eq:klimmicr}
N_M(\vec x,\vec p,t) = \sum\limits_{i=1}^N \delta^{(3)}(\vec x - \vec x_i(t)) \, 
\delta^{(3)}(\vec p - \vec p_i(t)) 
\end{equation}
introduced by Klimontovich \cite{Kl} for a classical $N$-particle system. 
I first will briefly outline in which context Klimontovich uses this function:
Aiming at a statistical description of a classical system of $N$ interacting particles 
the quantity (\ref{eq:klimmicr}) is introduced as the phase space number density of a 
specific microscopic realization of the macroscopic system. Assuming that the particles
interact via central two-body forces obtained from the potential 
$\Phi(\vert \vec x_i - \vec x_j \vert)$ one obtains Newton's equations for the 
coordinates of particle $i$:
\begin{equation}
  \label{eq:newton1}
\dot {\vec x}_i = {\vec p_i \over m} \,, \qquad 
\dot{\vec p}_i 
= -\vec \nabla_{x_i} \sum\limits_j \Phi(\vert \vec x_i - \vec x_j \vert)  \,.
\end{equation}
Combined with the equation of continuity in phase space 
$d N_M/ dt = 0$ one gets
\begin{equation}
  \label{eq:klimeq}
\partial_t N_M + {\vec p \over m} \vec \nabla_x N_M
- \vec \nabla_x \int \!\! d^3\!x'\,d^3\!p' 
\Phi(\vert \vec x - \vec x' \vert) N_M(\vec x',\vec p',t)
\vec \nabla_p N_M = 0  
\end{equation}
called the Klimontovich equation. 
Formally this equation resembles the Vlasov equation. However, there is an important
difference between the Klimontovich and the Vlasov equation: The latter is an 
approximate equation for the ensemble averaged one-particle distribution function $f$
(which can be obtained e.g.~from (\ref{eq:boltz}) by neglecting the collision term 
$I_{\rm coll}$). In contrast, no statistical information has entered (\ref{eq:klimeq}) 
yet. There, one microscopic realization of the $N$-particle system is considered. The
crucial step which involves statistics amounts to average (\ref{eq:klimeq}) over
all microscopic realizations which form the macroscopic state, i.e.~to consider an 
average weighted by the $N$-particle phase space distribution function. In this way
one deduces from (\ref{eq:klimeq}) an equation of motion for the ensemble averaged 
quantity $\bar N_M$ which coincides (up to normalization) with the 
one-particle phase space density $f$. Obviously in the resulting equation the average
over the product of two microscopic phase space densities appears (stemming from the 
two-body interaction part). From (\ref{eq:klimeq}) one can also deduce an equation of 
motion for such a product which in turn involves products of three microscopic phase 
space densities. In this way one generates a hierarchy of equations which is analogous
to the BBGKY hierarchy of $n$-particle distribution functions. It is the merit of
Klimontovich's method that the whole hierarchy can be deduced from one 
equation (\ref{eq:klimeq}). If in the ensemble average of (\ref{eq:klimeq}) the average
of the product of two microscopic phase space densities is approximated by the product
of averages then one obtains the Vlasov equation. In this case all possible correlations
are neglected. If such correlations are retained one can obtain --- using a proper 
approximation scheme --- e.g.~collision terms of 
Boltzmann type. To summarize, in Klimontovich's method 
the microscopic phase space density (\ref{eq:klimmicr}) is introduced to obtain 
(\ref{eq:klimeq}) from which all equations of motion for ensemble averaged products of
these densities can be deduced. Within proper approximation schemes the well-known 
(ensemble averaged) kinetic equations with and without collision terms can be obtained. 
Now one might turn the procedure around: Starting from (\ref{eq:klimeq}) one might
try to find a proper way to solve this equation. This can be achieved by the ansatz 
(\ref{eq:klimmicr}) which one might be tempted to call ``test particle ansatz''. (This,
however, is misleading.)
One obtains in this way the equations of motion (\ref{eq:newton1}). Solving these
equations e.g.~by a numerical simulation yields indeed {\em one} solution of 
(\ref{eq:klimeq}). This, however, is not
the whole story. One is actually interested in the ensemble averaged $n$-particle 
distribution functions. To obtain these one has to solve (\ref{eq:newton1}) for various
initial conditions which belong to the same macroscopic state. Calculating averages
and fluctuations of the various simulations (characterized by the initial conditions)
yields the quantities of interest. To compare this strategy to the test particle
ansatz discussed in the previous sections one has to realize that it is the purpose of 
the latter to yield
a practicably applicable method to solve the {\em ensemble averaged} Boltzmann equation
and not a microscopic equation. Thus one does not start with arbitrary initial
conditions for the test particle coordinates but instead with a large number of test 
particles such that the initial one-particle phase space distribution
is reasonably well reproduced by the test particle distribution.
A second difference concerns the collision
terms, i.e.~dissipative terms which are explicitly present in the Boltzmann equation 
(\ref{eq:boltz}) but not in the microscopic Klimontovich equation (\ref{eq:klimeq}).
While the equations of motion for
the test particles {\em between collisions} are --- for a classical system --- indeed 
identical to the microscopic equations (\ref{eq:newton1}) the test particles in addition
scatter on each other with the respective scattering cross sections which enter the 
Boltzmann collision terms. To summarize, in spite of the strong formal similarity
between the test particle ansatz (\ref{eq:testp1}) and Klimontovich's microscopic
phase space density (\ref{eq:klimmicr}) there are important conceptual differences in
the two approaches. This becomes even more pronounced if genuine quantum corrections
like the off-shell effects discussed above are included in the test particle approach.

\section{Summary}

I have presented a scheme to derive from the Kadanoff-Baym equations
a transport equation beyond the quasi-particle approximation for states with potentially
large width. I have argued that the particle number $N$ defined for the full quantum 
field
theory is not exactly conserved by the transport equation. Instead it can be shown that
the effective particle number $\tilde N$ is conserved. I have derived equations
of motion for the test particle coordinates from a test particle ansatz for 
$\tilde S^<$. Finally I have presented the modifications of the collision integrals
once they are used to describe collisions between the test particles. Once the transport
equation is solved by this ansatz one can calculate $dN/dt$ and find out to which amount
the conservation of $N$ is violated. This provides a test for the accuracy of the
gradient expansion. 

\vspace{0.2cm}
{\bf Acknowledgments:} I would like to thank the organizers and especially Michael 
Bonitz for providing the basis for an interesting and inspiring workshop.


\section*{References}
\begin{thebibliography}{99}

\bibitem{SL99} S.~Leupold, nucl-th/9909080, submitted to {\bf Nucl.~Phys.~A}.

\bibitem{Kl} Yu.~L.~Klimontovich, Sov.~Phys.~JETP, {\bf 6}, 753 (1958);
{\bf 35}, 920 (1972); The statistical theory of non-equilibrium 
processes in a plasma, Pergamon Pr., Oxford, 1967.

\end{thebibliography}
\end{document}